\pdfoutput=1
\documentclass[sigconf]{acmart}

\usepackage{fancyhdr}
\AtBeginDocument{%
    \addtolength{\footskip}{2.0\baselineskip}%
    \fancyfoot[L]{
    \providecommand\BibTeX{{%
    \normalfont B\kern-0.5em{\scshape i\kern-0.25em b}\kern-0.8em\TeX}}}%
}

\copyrightyear{2022}
\acmYear{2022}
\setcopyright{acmlicensed}\acmConference[AISec '22]{Proceedings of the 15th
ACM Workshop on Artificial Intelligence and Security}{November 11, 2022}{Los
Angeles, CA, USA}
\acmBooktitle{Proceedings of the 15th ACM Workshop on Artificial Intelligence
and Security (AISec '22), November 11, 2022, Los Angeles, CA, USA}
\acmPrice{15.00}
\acmDOI{10.1145/3560830.3563726}
\acmISBN{978-1-4503-9880-0/22/11}

\usepackage{array,booktabs}
\newcolumntype{P}[1]{>{\centering\arraybackslash}p{#1}}
\newcolumntype{M}[1]{>{\centering\arraybackslash}m{#1}}

\usepackage{pdfpages}

\begin{document}


\title[Hybrid Machine Learning Meta-Model]{Quo Vadis: Hybrid Machine Learning Meta-Model Based on Contextual and Behavioral Malware Representations}

\author{Dmitrijs Trizna}
\email{dtrizna@microsoft.com}
\affiliation{%
  \institution{Microsoft Corporation}
  \country{Prague, Czech Republic}
}

\renewcommand{\shortauthors}{Dmitrijs Trizna}

\begin{abstract}

We propose a hybrid machine learning architecture that simultaneously employs multiple deep learning models analyzing contextual and behavioral characteristics of Windows portable executable, producing a final prediction based on a decision from the meta-model. The detection heuristic in contemporary machine learning Windows malware classifiers is typically based on the static properties of the sample since dynamic analysis through virtualization is challenging for vast quantities of samples. To surpass this limitation, we employ a Windows kernel emulation that allows the acquisition of behavioral patterns across large corpora with minimal temporal and computational costs. We partner with a security vendor for a collection of more than 100k int-the-wild samples that resemble the contemporary threat landscape, containing raw PE files and filepaths of applications at the moment of execution. The acquired dataset is at least ten folds larger than reported in related works on behavioral malware analysis. Files in the training dataset are labeled by a professional threat intelligence team, utilizing manual and automated reverse engineering tools. We estimate the hybrid classifier's operational utility by collecting an out-of-sample test set three months later from the acquisition of the training set. We report an improved detection rate, above the capabilities of the current state-of-the-art model, especially under low false-positive requirements. Additionally, we uncover a meta-model's ability to identify malicious activity in both validation and test sets even if none of the individual models express enough confidence to mark the sample as malevolent. We conclude that the meta-model can learn patterns typical to malicious samples out of representation combinations produced by different analysis techniques. Furthermore, we publicly release pre-trained models and anonymized dataset of emulation reports. 

\end{abstract}



\maketitle

\section{Introduction}

Machine learning (ML) algorithms have become essential to malicious software (malware) detection in conventional cybersecurity intrusion prevention systems. Such systems can learn common patterns across a vast malware dataset, obtaining a predictive power to classify previously unseen malicious samples. However, there is evidence that contemporary state-of-the-art models lack epistemic capacity due to limited contextual and behavioral awareness \cite{kyadige2019learning} since they mostly rely on representations acquired from static properties of the executable files \cite{anderson2018ember, raff2017malconv}.

Human-produced malware analysis typically is based on static and dynamic properties of sample \cite{FOR610}. Static evaluation of malicious specimen provides readily available yet limited insights on its functionality, usually surpassed by a dynamic analysis through sample ``detonation'' in a controlled environment. However, the collection of Windows portable executable (PE) behavioral patterns through dynamic analysis sufficient for ML algorithms, especially if based on deep learning architectures, poses a significant challenge due to the computational burden of virtualization technology and the necessity to revert operating system setup from the contamination after malware detonation. 


We perform dynamic analysis with malware-oriented Windows kernel emulator \cite{SpeakEasy}, thus achieving high analysis rates compared to virtualization. Because of data heterogeneity, we consider a composite solution with multiple individual pre-trained modules and a meta-model rather than building a single feature vector with end-to-end trainable architecture. This architecture allows extending the modularity of the decision heuristic with minimal efforts by retraining only a meta-model. In the scope of this publication assessment of hybrid ML architecture relied on three distinct analysis techniques:

\begin{itemize}
    \itemsep0em
    \item contextual information in the form of a file path on a system at the moment of execution;
    \item sample behavior expressed as a sequence of Windows kernel API calls;
    \item static representations obtained from the Windows PE structure.
\end{itemize}

We expect further work on additional behavioral models like network or filesystem telemetry analysis. Therefore, we release an anonymized emulation dataset publicly. The lack of publication artifacts is a notorious drawback in any research and contributes to science's reproducibility crisis. Hence we disclose\footnote{\url{https://github.com/dtrizna/quo.vadis}} the source code and pre-trained PyTorch \cite{pytorch}, and scikit-learn \cite{saxe2017expose} models, as well as provide \texttt{scikit-learn}-like \cite{scikitlearn} API for our model adhering to widely adopted interface of a machine learning objects. To our knowledge, we are the first in the security research community to publish a model that incorporates a single decision heuristic based on (a) contextual, (b) static, and (c) dynamic properties of the PE file. 

We collect an out-of-sample dataset three months after model training. We acknowledge that malware classification based on hybrid representations of software yields improved detection performance and reduced false-positive rates against the evolving nature of malevolent logic compared to any individual method capabilities. Additionally, we evaluate adversarial robustness of our model based on section injection attack defined by Demetrio et al. \ref{fig:gamma_structure}, and report evasion rates in a variable modularity configuration.

This article is structured as follows - Section \ref{sec:rel_work} reviews related work, Section \ref{sec:dataset} describes dataset and its collection specifics, Section \ref{sec:arch} covers architecture of hybrid architecture and data preprocessing, Section \ref{sec:results} reports performance and adversarial robustness of our model, Sections \ref{sec:conclusions} draws conclusions from empirical observations and outlines future work options.

\section{Related Work}\label{sec:rel_work}

Since the idea of malware detection using ML techniques was introduced by Schultz et al. \cite{first_paper}, the field in research and industry has grown significantly. The prevailing part of ML malware classification is based on static analysis techniques, ML heuristics are applied to representations acquired from fixed properties of malware files \cite{sihwail2018survey, ucci2019survey}. One of the first neural network applications for malware classification was shown by Raff et al.~\cite{raff2017malconv} who presented a \emph{MalConv} model - a "featureless" Deep Neural Network (DNN) that reads raw bytes of executable and proceeds with embeddings and one-dimensional convolution.

Gradient Boosted Decision Tree (GBDT) models achieve notable success in malware classification, specifically the approach introduced by Anderson et al. \cite{anderson2018ember}. Their work is meant to provide a benchmark dataset based on specific, pre-extracted properties from malware files, thus the title: \emph{EMBER} (Endgame Malware BEnchmark for Research). At the same time, the paper includes an evaluation of LightGBM \cite{NIPS2017_lightGBM} model performance, describing an approach that employs a clever feature engineering phase. Ember representations incorporate domain knowledge into many effective static characteristics of PE files, becoming a \emph{de facto} standard for static feature extraction from PE files in modern malware classification research. An interesting approach is shared by Rudd et al. \cite{rudd2019aloha}, which utilizes Ember representation vectors to train a feed-forward neural network (FFNN).

ML algorithms are proven fruitful by utilizing dynamic analysis telemetry from malware "detonation" in a sandbox. Generally, dynamic PE analysis methods use API call telemetry to represent PE activity. Rosenberg et al. \cite{rosenberg2020attack_on_api_calls} construct a one-hot encoded vector out of encountered API calls. This approach is the simplest possible and ignores API sequences. Kolosnjaji et al. \cite{kolosnjaji2016sequences} showed that it is possible to perform API call processing to preserve sequential information. Another example is described by Yen et al. \cite{yen2019integration}, who obtain a behavioral representation based on API call frequency.

The hybrid analysis allows to surpass the limitations of each malware analysis method, and several security research groups provided insights on utilizing the hybrid approach with ML algorithms. For example, Shijo and Salim \cite{SHIJO2015804} document a way to construct a feature vector leveraging data from static and dynamic analysis techniques and processed by a single ML model. On the contrary, Ma et al. \cite{ma2016hybrid} use an ensemble of different classifiers to perform a hybrid analysis, similarly to modeling techniques proposed in our work, building separate feature sets from static and dynamic analysis telemetry.

Observing malware in a sandbox is costly in terms of required computational resources and execution time. Therefore, it is hard to collect dynamic analysis telemetry in quantities beneficial for most ML algorithms, especially based on deep learning architectures, with conclusions that generalize well across the true distribution of malicious sample properties. For instance, Shijo and Salim \cite{SHIJO2015804} evaluate their technique on 997 virus and 490 clean files, Ma et al. \cite{ma2016hybrid} use 282 samples, Kolosnjaji et al. \cite{kolosnjaji2016sequences} have a dataset of 4753 executables, Yen et al. \cite{yen2019integration} use 4519 files.

Emulators do not require to mobilize full-fledged operating system operations, as they allow getting vast amounts of telemetry reasonably fast, without the need for virtualization infrastructure. Therefore, the utilization of emulators as a telemetry source of ML models for dynamic malware analysis research is promising yet not commonly adopted. To our knowledge, the first occurrence of emulator utilization for system call collection was reported by Athiwaratkun and Stokes in 2017 \cite{emulation2017lstm}. Their model resembles recurrent schemes used in Natural Language Processing (NLP). This work is further developed by Agrawal et al. \cite{agrawal2018robust} who present a similar architecture adopted for arbitrary long API call sequences acquired with the help of an emulator. Mandiant's data science team performs promising research with emulation-based dynamic analysis. Specifically, Li et al. \cite{Li2021CAMLISmandiant} provides an extended abstract that reports utilization of emulator \cite{SpeakEasy} for hybrid analysis with architecture similar to ours.

The sparsity of work on emulation-based behavior analysis is due to their limitations. Emulation is an abstraction on top of the operating system where the emulator runs, and no direct interaction with hardware happens. Theoretically, the perfect emulator could spoof the logic behind any system call. Nonetheless, kernels like Windows NT incorporate massive functionality, yielding the implausible achievement of one-to-one replicas. Hence real-world emulators implement only a subset of all possible kernel manipulations, and sophisticated malware samples can identify a limited emulation environment, preventing detailed behavior analysis.

We report a detailed error rate and data diversity comparison to related virtualization-based work datasets in Section \ref{sec:emulation}. We argue that modern Windows kernel emulation has a vast potential in ML malware detectors based on empirical evidence. Emulation reports produce rich and diverse telemetry, bypassing static analysis limitations. It contains a sequence of kernel API calls invoked by the executable and describes manipulations with files or registry entries and attempted network communications.

Significant limitation of a modern static classifiers is weak defense against adversarial malware samples \cite{advbinaries, anderson2018rl, demetrio2021gamma, kucuk2020genetic, song2021mabmalware}. Therefore, we consider adversarial robustness of our model against contemporary evasion attacks. We evaluate model performance against attack based on genetic algorithm with a black-box threat model, specifically, GAMMA (Genetic Adversarial Machine learning Malware Attack) as formulated by Demetrio et al. \cite{demetrio2021gamma}.

\section{Dataset}\label{sec:dataset}

\begin{figure*}[ht]
    \centering
    \includegraphics[width=0.9\textwidth]{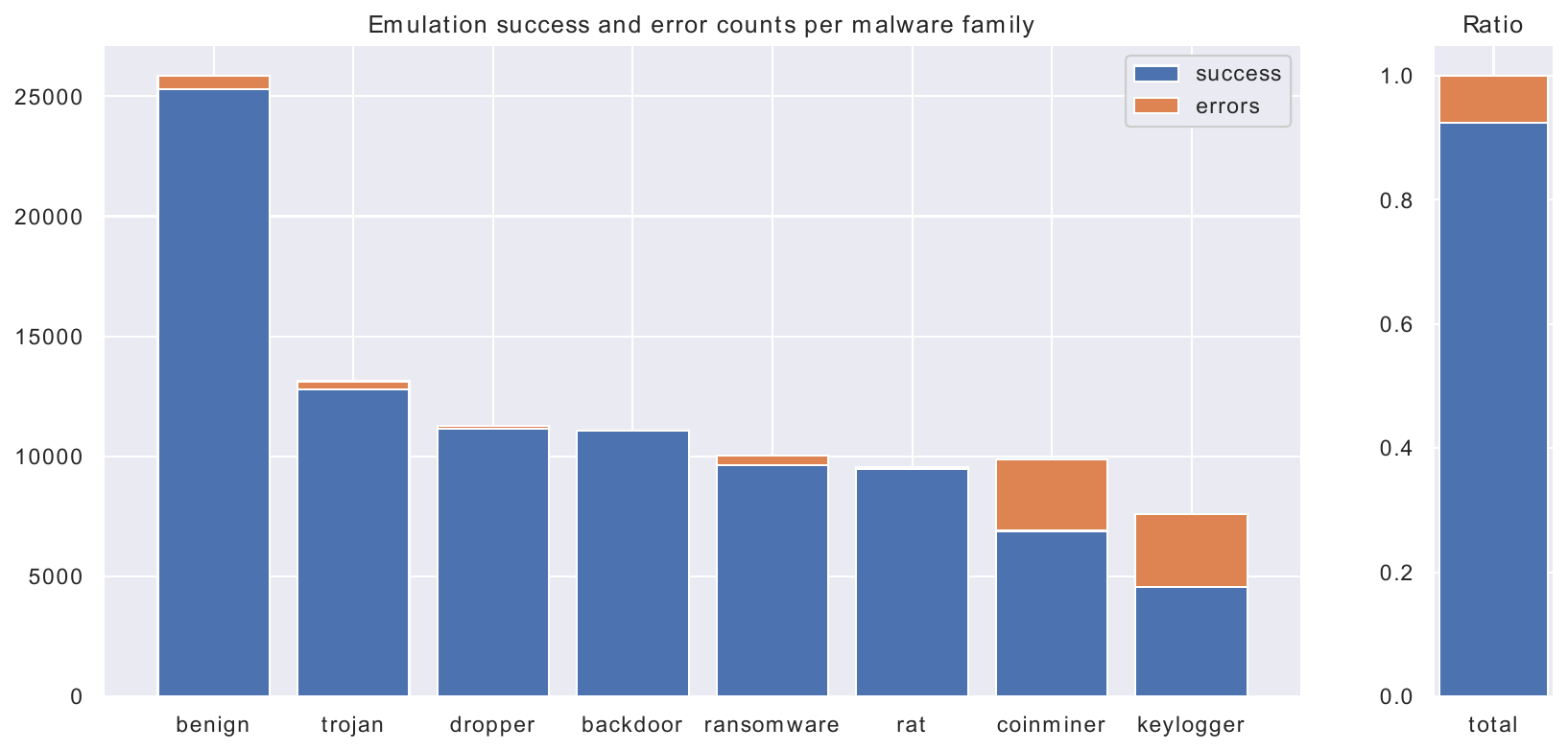}
    \caption{PE emulation error distribution across malware families in in-sample training and validation sets.}
    \Description{Figure Describes an error ratio across every individual malware family and in a total dataset. Erroneous were considered emulation reports that did not produce any API call telemetry, and exited with invalid memory read or write operations. Successful were considered reports that produced $>=1$ more API calls.}
    \label{fig:emulation_err}
\end{figure*}

The functionality of a hybrid solution presented in this work is based on input data consisting of both (a) armed PE files suitable for dynamic analysis \textit{and} (b) contextual filepath information. The necessity to acquire contextual data yields impossible the utilization of public data collections since for \textit{every} data sample we need to possess both raw PE bytes \textit{and} filepath data on an in-the-wild system. 

To the best of our knowledge, none of the known datasets provide contextual information about PE samples with filepath values at the moment of file execution. For example, Kyadige and Rudd et al. \cite{kyadige2019learning} rely on a proprietary Sophos' threat intelligence feed and do not release their dataset publicly. The private nature of contextual data is understandable since such telemetry would contain sensitive components, like directories on personal computers.

Therefore, we partner with an undisclosed security vendor for a vast dataset collection, containing both raw PE files and filepaths of samples from personal customer systems that would resemble an up-to-date threat landscape. Data is treated with respect to the privacy policy accepted by all customers. Therefore, we do not publicly release the file path and raw PE dataset. Sensitive data components like usernames or custom environment variables are from telemetry during the pre-processing stage and have not been resembled within model parameters or emulation reports.

\subsection{Dataset structure}

We collect the dataset in two sessions. The first session forms the foundation of our analysis, consisting of \textbf{98 966} samples, \texttt{329 GB} of raw PE bytes. $80\%$ of this corpus is used as a fixed training set, and $20\%$ form an in-sample validation set. We pre-train models and investigate our hybrid solution configuration using this data.

The second dataset acquisition session occurred \textit{three months later}, forming an out-of-sample test set from \textbf{27 500} samples, about \texttt{100 GB} of data. This corpus is used to evaluate the real-world utility of the hybrid model and investigate model behavior on the evolved malevolent landscape. 

The PE files in the dataset are tagged by a professional threat intelligence team, utilizing manual and automated reverse engineering tools operated by the malware analysts. The dataset spans seven malware families and benignware, with detailed distribution parameters described in Table \ref{table:dataset}. All labels except ``\textit{Clean}'' represent malicious files. Therefore, we collected relatively more ``Clean'' samples to balance malicious and benign labels in the dataset.

\begin{table}[ht]
    \centering
    \caption{Dataset structure and size.}
    \label{table:dataset}
    \begin{tabular}{|c|c|c|c|c|}
        \hline
        & \multicolumn{2}{c|}{Train \& valid. sets} & \multicolumn{2}{c|}{Test set} \\
        \hline
        \textbf{File label} & \textbf{Size (Gb)} & \textbf{Counts} & \textbf{Size (Gb)} & \textbf{Counts} \\
        \hline
        Backdoor & 30.0 & 11089 & 7.4 & 2500 \\
        \hline
        \textit{Clean} & 127.0 & 26061 & 47.0 & 10000 \\
        \hline
        Coinminer & 46.0 & 10044  & 11.0 & 2500\\
        \hline
        Dropper & 36.0 & 11275  & 9.0 & 2500\\
        \hline
        Keylogger & 34.0 & 7817  & 9.8 & 2500\\
        \hline
        Ransomw. & 14.0 & 10014  & 4.6 & 2500\\
        \hline
        RAT & 5.5 & 9537  & 2.5 & 2500\\
        \hline
        Trojan & 40.0 & 13128  & 7.1 & 2500\\
        \hline
        \textbf{Total} & 329 & 98966 & 98 & 27500 \\
        \hline
    \end{tabular}
\end{table}


We focus on 32-bit (x86) images and deliberately skip the collection of 64-bit (x64) images to maintain homogeneity and label balance of the dataset. The dataset is formed out of executables (\texttt{.exe}), and we intentionally omit library PE files (\texttt{.dll}).

\subsection{Sample emulation}\label{sec:emulation}

All the samples represented in Table \ref{table:dataset} were processed with a Windows kernel emulator. We utilize Speakeasy \cite{SpeakEasy} Python-based emulator released and actively maintained by Mandiant under MIT license. The Speakeasy version used in our tests is 1.5.9. It relies on QEMU \cite{qemu} CPU emulation framework. 

We obtained \textbf{108204} successful emulation reports, emulating 90857 samples from training and validation sets, and 17347 samples in the test set. Unfortunately, some sample emulations were erroneous, primarily due to an invalid memory read of write assembly instructions. However, another common reason for emulation errors is a call of unsupported API function or anti-debugging techniques. Figure \ref{fig:emulation_err} shows the error rate across different malware families. 

Dataset processing with an emulator had only minimal computational overhead. Overall report emulation time mode is only $0.25s$, with a $75\%$ quantile of $7.63s$. Additionally, emulation can be distributed across separate CPU or GPU cores, yielding multi-threaded processing, a significant benefit over a sandbox environment. These observations prove that emulation possesses a real-world utilization for behavioral analysis, for instance, considering samples submitted to a cloud back-end by an endpoint detection and response (EDR) agent.

\begin{table*}[ht]
    \centering
    \caption{Dataset diversity preserved and in-sample validation set's F1-score based on choice of top API calls.}
    \label{tab:apicall_stats}
    \begin{tabular}{|c|c|c|c|c|c|c|c|c|c|}
        \hline
        Top API calls & 100 & 150 & 200 & 300 & 400 & 500 & 600 & 700 \\
        \hline
        Dataset \% & 95.53 & 97.67 & 98.73 & 99.48 & 99.74 & 99.85 & 99.91 & 99.94 \\
        \hline
        Val. F1-score & 0.9707 & 0.9712 & 0.9725 & 0.9740 & 0.9752 & 0.9747 & \textbf{0.9759} & 0.9754 \\
        \hline
    \end{tabular}
\end{table*}




We observe that emulation forms a diverse and rich telemetry. For instance, we acquire \textbf{2822} unique API calls within training and validation dataset reports. This behavior is significantly more heterogeneous than in related work datasets - Athiwaratkun and Stokes \cite{emulation2017lstm} have a total of 114 unique API calls, Kolosnjaji et al. \cite{kolosnjaji2016sequences} report 60 unique API calls, Yen et al. \cite{yen2019integration} have 286 different API calls, Rosenberg et al. \cite{rosenberg2020attack_on_api_calls} have 314 individual API calls. Partially such observation can be described by the larger volume of our dataset since the number of unique calls positively correlates with the number of samples. However, we emphasize this as evidence of emulation technique efficiency, which allows to acquire wide set of behavioral patterns from within PE samples.


\section{System Architecture}\label{sec:arch}

\begin{figure*}[t]
    \centering
    \includegraphics[width=0.9\textwidth]{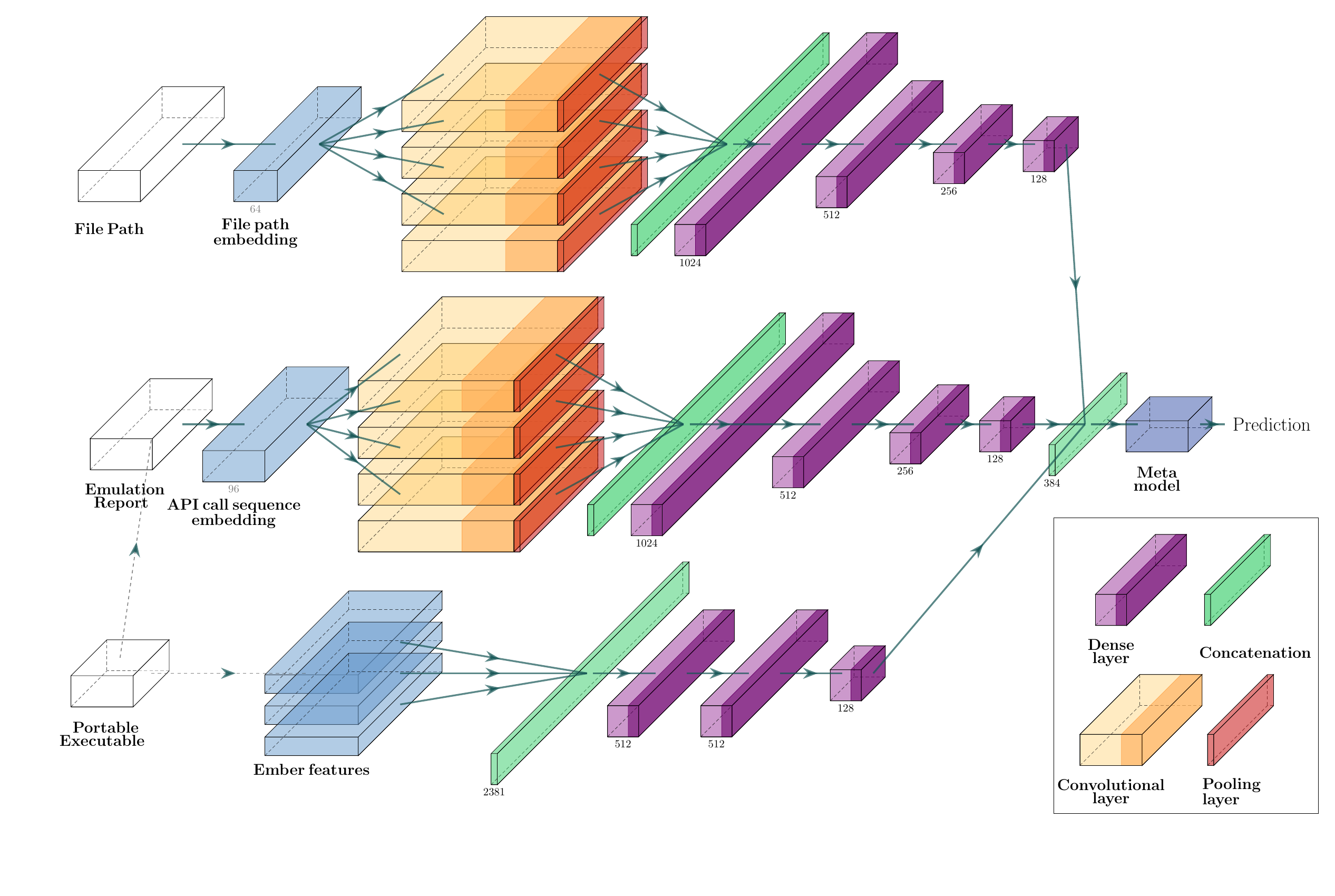}
    \caption{General view of hybrid model architecture with three separate modules.}
    \Description{Hybrid model consists of three independent, pre-trained units, each used to evaluate the distinct property of the PE file, forming an early fusion pass $\phi(x)$. Filepath and emulated API call models rely on input embeddings, and four 1D convolutional layers passed to FFNN. Ember produces custom features and uses them as input for the FFNN model. Furthermore, all three model output is used to train a meta-model that produces a final prediction.}
    \label{fig:composite_arch}
\end{figure*}

A general overview of the hybrid model architecture is visualized in Figure \ref{fig:composite_arch}. The composite architecture consists of multiple independent models $\phi$, that are ``fused'' together for a final decision produced by a meta-model $\psi$. Three early fusion models $\phi$ are:

\begin{itemize}
    \itemsep0em
    \item file-path 1D convolutional neural network (CNN), $\phi_{fp}$
    \item emulated API call sequence 1D CNN, $\phi_{api}$
    \item FFNN model processing Ember feature vector, $\phi_{emb}$
\end{itemize}

Each early fusion model reports a $128$-dimensional vector of representations acquired from input data. Furthermore, all three models' outputs are concatenated together forming an $384$-dimensional vector. Therefore, given an input sample $x$, consisting of raw PE as bytes and its filepath as string, early fusion pass collectively is denoted as:

\begin{equation}
    \label{eq:early_fusion}
\phi(x) = [\phi_{fp}, \phi_{api}, \phi_{emb}] \in \mathbb{R}^{384}
\end{equation}

The intermediate vector $\phi(x)$ is passed to a meta-model $\psi$, which produces the final prediction:

\begin{equation}
    \label{eq:late_fusion}
    \hat{y} = \psi(\phi(x)) \in [0,1]
\end{equation}

All early fusion networks are pre-trained separately with detailed definition of model configurations and training process in Section \ref{sec:exp}. We want to emphasize that decision to construct a modular system with multiple individually pre-trained components instead of building a single end-to-end trainable architecture is deliberate. First of all, it is shown by Yang et al. \cite{yang2019composite} that composite neural networks with a high probability surpass the performance of individual pre-trained components. 

However, the main reason is the vast potential for expanding a hybrid decision heuristic with complementary modules by only retraining a meta-model. Malicious activity classification problem relies on highly heterogeneous information sources, beyond raw PE bytes, and such architecture preserves ability on adding heuristics that rely on system logging. At the moment of this publication we already incorporate filepath information, that can be acquired, for instance, from Sysmon\footnote{\url{https://docs.microsoft.com/en-us/sysinternals/downloads/sysmon}} telemetry. However, knowledge from within Sysmon data or Speakeasy reports can be extracted further. 

Additionally, we acknowledge that filepaths alone do not provide enough knowledge \emph{per se} to classify a file as malicious or benign. Filepaths, however, often give insights on the ``fitness'' of a file to a usual operating system operation. Existing research base shows, that supplementing static PE detection analysis with additional information in form of filepath improves detection performance \cite{kyadige2019learning}. 


\subsection{API Call Preprocessing}

To acquire a numeric value of API call sequences, we select the top most common calls based on variable vocabulary size $V$. Preserved API calls are label-encoded, and calls that are not part of the vocabulary are replaced with a dedicated label. The final sequence is truncated or padded using a padding label to a fixed length $N$.

Table \ref{tab:apicall_stats} shows the statistics behind dataset diversity and respective model performance. Even though 100 most common calls contain more than 95\% API calls within a dataset, experiments show that the model still benefits from relatively large vocabulary size values, so we have chosen $V=600$ for our final configuration. This observation might be explained by the distribution of API calls per sample. Verbose executables with hundreds of calls bias system call frequency, whereas executables with modest API sequences perform more unique function combinations.

\subsection{Path Preprocessing}

The first part of path pre-processing includes path normalization since some parts of filepath semantics have variability that is irrelevant for security analysis through the deep learning model. These include specific drive letters or network location if a universal naming convention (UNC) format is used, as well as individual usernames. Therefore, during normalization, we introduced universal placeholders for those path components as presented below:

\begin{verbatim}
    [drive]\users\[user]\desktop\04-ca\8853.vbs
    [drive]\users\[user]\appdata\local\file.tmp
    [net]\company\priv\timesheets\april2021.xlsm
\end{verbatim}

Additionally, it is necessary to parse Windows environment variables to resemble the actual filepath rather than the environment alias used as a variable name. Therefore, we built a variable map consisting of about 30 environment variables that represent specific paths on a system and are used across contemporary and legacy Windows systems. Some examples of a variable map are presented below:

\begin{verbatim}
    r"%systemdrive%": r"[drive]", 
    r"%systemroot%": r"[drive]\windows",
    r"%userprofile%": r"[drive]\users\[user]"
    ...
\end{verbatim}

We perform the encoding of unique letters against the UTF-8 character set. A similar approach was used by Saxe et al. \cite{saxe2017expose} when evaluating URL maliciousness, and by Kyadige and Rudd et al. \cite{kyadige2019learning} on a filepath data, using 100 and 150 most frequent UTF-8 bytes, respectively. Rare characters below a frequency threshold are discarded and replaced by a single dedicated label.

\subsection{Early Fusion Model Architectures}
\label{sec:exp}

As mentioned in Section \ref{sec:dataset}, we cannot rely on publicly released malware collections since the model requires contextual information in the form of filepaths at the moment of execution, which is not available publicly. Therefore, to preserve the ability of assessment model performance in comparison to existing research, we can rely only on malware classification models released with pre-trained parameters by other research groups, with further evaluation of those on our dataset. Unfortunately, none of hybrid or dynamic analysis publications \cite{rosenberg2020attack_on_api_calls, kolosnjaji2016sequences, yen2019integration, SHIJO2015804, ma2016hybrid} provide such artifacts.

Luckily, multiple static analysis publications were accompanied by artifacts in the form of the code repositories \cite{raff2017malconv, anderson2018ember, rudd2019aloha}. For instance, it is possible directly use Ember LightGBM \cite{NIPS2017_lightGBM} model, pre-trained on Ember dataset \cite{anderson2018ember} as released in 2019 by Endgame\footnote{\url{https://github.com/endgameinc/malware_evasion_competition}}. However, we do not include this model in our composite solution since the decision tree model does not learn representations that can be used by the meta-model, providing only final prediction in scalar form. We still rely on Ember feature extraction scheme \cite{anderson2018ember}, but use a FFNN as published by Rudd et al. \cite{rudd2019aloha} with three hidden layers or $512$, $512$, and $128$ hidden neurons respectively, all using ELU \cite{Clevert2015elu} non-linearity, with layer normalization \cite{Ba2016layernorm} and dropout \cite{dropout} rate $p=0.05$. We retrained FFNN for $200$ epochs on 600k feature vectors from Ember training set \cite{anderson2018ember} and 72k samples from our training set. 

\begin{figure*}[ht]
    \centering
    \includegraphics[width=0.8\textwidth]{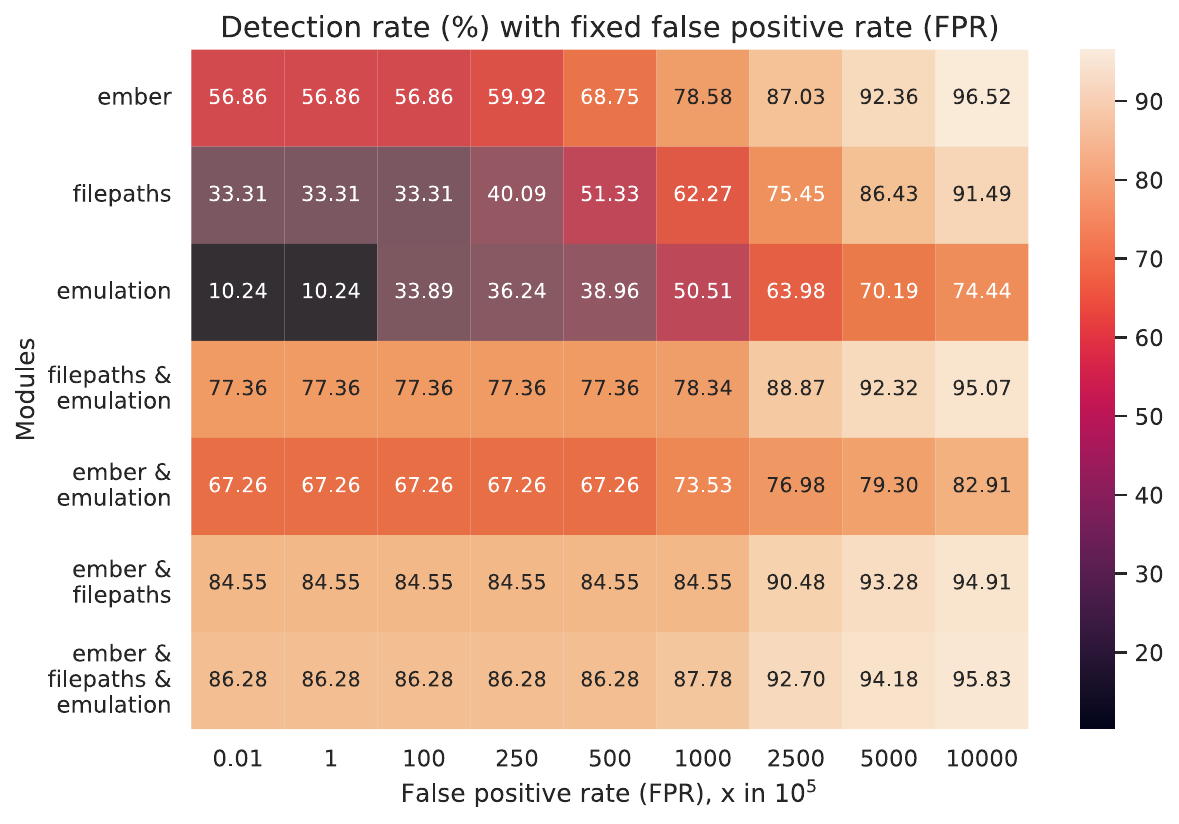}
    \caption{Detection rate (\%) on out-of-sample test set with fixed false positive rate based on different combinations of enabled modules in hybrid solution.}
    \Description{Heatmap of Detection rates given fixed false positive rates provides visual intuition on how independent model units complement each other. Color marking allows to clearly see a detection rate gap between one different module combinations.}
    \label{fig:rate_heatmap}
\end{figure*}

The analysis of file path and API call sequences can be formulated as a related optimization problem, namely the classification of a 1-dimensional (1D) sequence. We utilize similar neural architecture in both models influenced by Kyadige and Rudd et al. \cite{kyadige2019learning}, namely embedding layer with a 1D convolutional neural network (CNN) for representation extraction and a fully connected neural network learning classifiers function. We are aware of multiple choices to model sequence classification problems with alternate architectures, such as recurrent neural networks (RNN) \cite{lstm, gru}. However, as shown in related work on API call sequence classification, both model architectures report similar performance \cite{kolosnjaji2016sequences, rosenberg2020attack_on_api_calls}, yet it is shown that 1D CNN is significantly less computationally demanding \cite{cnnvsrnn}.

Encoded input vector $x$ with fixed length $N$ is provided to embedding layer with dimensions $H$ and vocabulary size $V$. These parameters are subject to hyperparameter optimization. The optimal values for file path model obtained by hyperparameter optimization on validation set are: input vector $x_{fp}$ length $N=100$, embedding dimension $H=64$, vocabulary size $V=150$. Respective values for emulated API call sequence model are: input vector $x_{em}$ length $N=150$, embedding dimension $H=96$, and vocabulary size $V=600$. The vocabulary of the file path model is formed out of the most common UTF-8 bytes, and for the API call sequences model, the most common system calls are selected. Both vocabularies are enriched with two labels used for padding and rare characters.

The output of the embedding layer is passed to four separate 1D convolution layers with kernel sizes of $2, 3, 4$, and $5$ characters, and the number of output channels $C=128$. With lower $C$ values model underperforms. For instance, having $C=64$ file path's module validation set F1-score is as low as $0.962$, while with $C \in \{100,128,160\}$ scores plateau around $0.966$.

The output of all four convolution layers is concatenated to a vector of size $4\times~C$ and passed to a FFNN with four hidden layers holding 1024, 512, 256, and 128 neurons. Hidden layers of FFNN are activated using rectified linear unit (ReLU) \cite{agarap2019deep}. During the pre-training phase we use the final layer with sigmoid activation, which is discarded when composite heuristic is used. Batch normalization \cite{ioffe2015batch} is applied to hidden layers of FFNN before the ReLU activation. Additionally, to prevent overfitting, dropout \cite{dropout} with a $p=0.5$ rate is applied. 

All early fusion networks are fitted using binary cross-entropy loss function:

$$ L(x, y; \theta) = -y \log(\phi(x; \theta)) + (1-y)\log(1-\phi(x;\theta)) .$$

$\phi(x;\theta)$ denotes function approximated by deep learning model given parameters $\theta$, and $y \in \{0,1\}$ are the ground-truth labels. Optimization is performed using Adam optimizer \cite{kingma2017adam} with $0.001$ learning rate and fixed batch size of $1024$ samples. We constructed both 1D convolutional networks, Ember FFNN, and training routine using PyTorch \cite{pytorch} deep learning library.

\subsection{Meta-Model}


\begin{table*}[ht]
    \centering
    \caption{In-sample validation set metrics assessed against various meta-model architectures.}
    \label{tab:late_modules}
    \begin{tabular}{|P{3cm}|c|c|c|c|c|P{2cm}|}
        \hline
        Model & AUC & F1-score & Recall & Precision & Accuracy & Convergence time \\
        \hline
        LR & \textbf{0.9987} & 0.9898 & 
                        0.9876 & 0.9920 & 
                        0.9853 & 2.52 s \\
        GBDT & 0.9986 & 0.9864 & 
                        0.9867 & 0.9862 & 
                        0.9803 & 34.92 s \\
        FFNN, 2 layers & 0.9973 & 0.9903 & 
                    0.9884 & 0.9923 & 
                    0.9860 & 14.30 s \\
        FFNN, 3 layers & 0.9965 & 0.9900 & 
                    0.9882 & 0.9918 & 
                    0.9855 & 24.52 s \\
        FFNN, 4 layers & 0.9957 & \textbf{0.9904} & 
                    0.9881 & \textbf{0.9926} & 
                    \textbf{0.9861} & 57.48 s \\
        FFNN, 5 layers & 0.9954 & 0.9901 & 
                    \textbf{0.9889} & 0.9913 & 
                    0.9857 & 36.73 s \\
        
        \hline
    \end{tabular}
\end{table*}

The output of early fusion models $\phi(x)$ is used to train the meta-model $\psi$. Three different architectures types were evaluated, Logistic Regression and FFNN were implemented using \texttt{scikit-learn} library \cite{scikitlearn}, and gradient boosted decision tree classifier based on \texttt{xgboost} \cite{xgboost} implementation.

Since meta-model $\psi$ performs a relatively complex non-linear mapping $[0,1]^{384} \rightarrow [0,1]$, based on performance metrics from Table \ref{tab:late_modules} we conclude that the fused classification surface is not smooth and presents combinations that utilizes representations from all three feature extraction methods for final decision, which simple models like Logistic Regression are not able to learn. We selected a four layer FFNN with $384, 128, 64,$ and $16$ neurons as a meta-model for our final evaluations since it has close to optimal scores.

\section{Experiments and Results}\label{sec:results}


\subsection{Detection rate analysis}

Experiments show that simultaneous utilization of static, dynamic, and contextual information yields significantly better detection rates than individual model performance, especially under low false positive requirements. Such demands are commonly expressed toward machine learning solutions in the security industry. Solutions that do not match low false positive needs are often not allowed to produce alerts for human analysts \cite{raff200fp}.

Detection rates (\%) given fixed false-positive rate (FPR) for the out-of-sample test set are visualised in Figure \ref{fig:rate_heatmap}. For instance, setting an alert threshold with FPR of only 100 misclassifications in $10^{5}$ cases, individual model detection rates are $56.86\%$, $33.31\%$, and $33.89\%$ for Ember FFNN, filepath, and emulation models respectively. However, by combining representations learned from all three processing techniques, the hybrid solution can correctly classify $86.28\%$ from all samples in the test set collected three months after training.

Surprising observations produce filepath and emulation models. Both models individually perform relatively poorly, especially if compared to ember FFNN. A potential explanation behind this observation is the ember FFNN training set consisting of 600k feature vectors from the original Ember publication \cite{anderson2018ember}. Such training corpora produce a much better generalization of "true" malicious PE distribution than our 100k samples reflecting threat landscape in a specific time window. 

However, under low false-positive requirements, just a combination of filepath and emulation model, omitting static analysis, outperforms the state-of-the-art Ember feature extraction scheme trained on a much broader dataset, with detection rates $77.36\%$ versus $55.86\%$ given FPR of one sample in $10^5$. 

\begin{table}[t]
    \centering
    \caption{Hybrid solution final metrics on validation and test sets with enabled Ember FFNN, API call sequence, and filepath modules using decision threshold of meta-model $0.98$.}
    \label{tab:final_metrics}
    \begin{tabular}{|c|P{1.5cm}|P{1.5cm}|P{1.5cm}|}
        \hline
        \textbf{Metric} & \textbf{Valid. set} & \textbf{Test set} \\
        \hline
        F1-score & 0.9900 & 0.9483 \\
        \hline
        Recall &  0.9865 & 0.9167 \\
        \hline
        Precision & 0.9934 & 0.9822 \\
        \hline
        Accuracy & 0.9855 & 0.9459 \\
        \hline
        AUC & 0.9847 & 0.9485 \\
        \hline
    \end{tabular}
\end{table}

Moreover, combining both models result in a detection rate above the cumulative capabilities of individual models, highlighting the hybrid meta-model's superiority over narrow solutions even more. For example, while individual filepath and emulation models detect only $33.31\%$ and $10.24\%$ of samples with FPR of one false alert in $10^5$, a combination of them produces a $77.36\%$ detection rate.

This observation holds across both in-sample validation and out-of-sample test sets, collected from divergent systems and in different time frames, allowing us to conclude that this is a general attribute of a hybrid detection heuristic with meta-model rather than an artifact of a specific dataset. Values for in-sample validation set given one misclassification in $10^5$ cases are detection rates of $34.46\%$, $13.52\%$, and $97.25\%$ for filepath, emulation, and combined heuristics, respectively.

This observation allows to conclude that the meta-model can learn patterns typical to malicious samples out of representation combinations produced by different analysis techniques, like combining a specific API call sequence and filepath n-gram. Each of these representations separately does not produce enough evidence to classify the sample as malicious since it also occurs in benign applications. Therefore, detection happens only by lifting false-positive requirements when both benign and malicious samples are flagged. However, a combination of representations from both filepaths and API calls allows for the meta-model to build a decision boundary in 384-dimensional space to segregate such cases, thus yielding detection rates of more than $40\%$.

As a result, we see that composite utilization of static, dynamic, and contextual data addresses independent method weaknesses, allowing minimization of FPR and false-negative rates (FNR). 

F1-score, Precision, Recall, Accuracy, and AUC scores on all sets are reported in Table \ref{tab:final_metrics} with meta-model decision threshold $0.98$ that resembled a FPR $\approx 0.25\%$ on validation set. While reported results on an in-sample validation set allow concluding that model has little to no overfitting, we still observe a decrease of an out-of-sample test set F1 and AUC scores by $\approx 4-4.5\%$. A drop in detection scores happens despite the same ratio of malware families, and we assume this phenomenon's causality arises from the evolving nature of malevolent logic.


\subsection{Adversarial robustness evaluation}

A large portion of adversarial malware research does not provide artifacts for reproducibility, reviewing the only conceptual description of attack methodology. The creation of adversarial malware samples is a subtle process with a high degree of nested dependencies and configuration variability. A direct replica of the attack without research artifacts like specific algorithm implementation or code sample is highly improbable. Luckily, attacks described by Demetrio et al.~\cite{demetrio2021gamma} are represented in accompanied \texttt{secml\_malware} library \cite{demetrio2021secmlmalware} focused on adversarial robustness evaluation of Windows malware classifiers. 

\begin{figure}
  \begin{center}
    \includegraphics[width=0.45\textwidth]{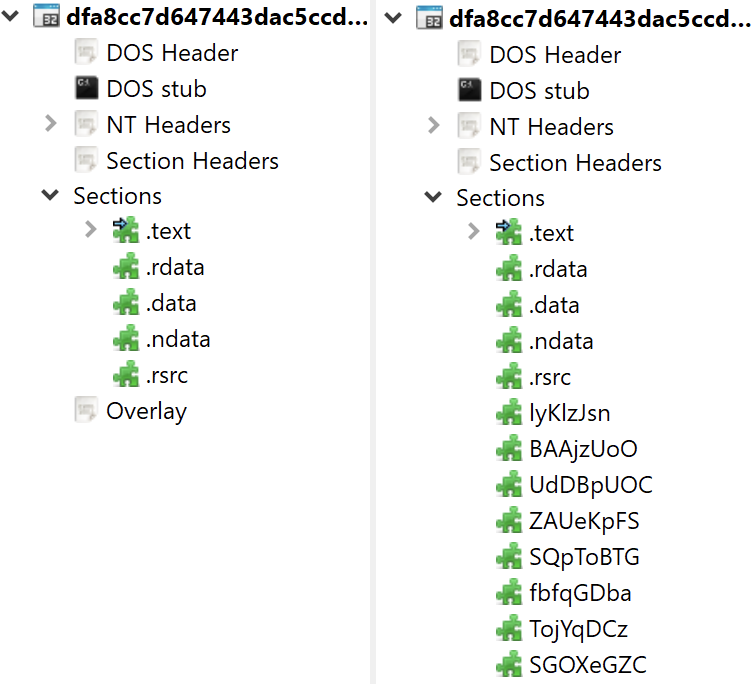}
  \end{center}
  \caption{Structure of PE file before (left) and after (right) GAMMA attack.}
  \label{fig:gamma_structure}
\end{figure}

We generate adversarial samples using section injection with GAMMA algorithm \cite{demetrio2021gamma}. Files after attack have additional PE sections as presented in Figure \ref{fig:gamma_structure}. We configure the algorithm to use \texttt{.rdata} and \texttt{.data} sections from legitimate software which is represented by a dozen of default Microsoft applications, for instance, \texttt{notepad.exe}, \texttt{calc.exe}, and \texttt{nslookup.exe}. Attack configuration is available in accompanying repository\footnote{\url{https://github.com/dtrizna/quo.vadis/blob/main/evaluation/adversarial/secml_malware/gamma_run.py}}.

GAMMA needs an \textit{oracle} to evaluate modified samples, and we select the GBDT model supported by \texttt{secml\_malware} to replicate attacks specified in original work on GAMMA \cite{demetrio2021gamma}. We launch an attack only against successfully detected malware samples by Ember module in the validation set for each target, in total $12604$ samples. Since the GAMMA algorithm performs multiple queries of intermediate sample versions during the attack, processing the whole corpus against takes roughly four days. Therefore, our options to evaluate different attack configurations are limited by resource constraints. However, we form an adversarial corpus using a variable number of injected sections, namely 5, 10, and 15 sections.

\begin{table*}[t]
    \centering
    \caption{Adversarial sample emulation statistics.}
    \begin{tabular}{|P{2.5cm}|P{2.5cm}|P{2.5cm}|P{2.5cm}|}
        \hline
        Target & Ember GBDT (15 sections) & Ember GBDT (10 sections) & Ember GBDT (5 sections) \\
        \hline
        Total samples & 8896 & 8995 & 9020 \\
        Successful & \textbf{5399} & 5438 & 5464 \\
        Errors & 3497 & 3557 & 3556 \\
        Success rate & 60.69\% & 60.46\% & 60.58\% \\
         \hline
    \end{tabular}
    \label{tab:adv_emulation_stats}
\end{table*}

\begin{table*}[ht]
    \centering
    \caption{Absolute count of evasive samples and evasion rates for both original and adversarial \textbf{5399} malware images forming an functional adversarial set after GAMMA attack with 15 section injection.}
    \begin{tabular}{|P{3cm}|P{2.5cm}|P{2.5cm}|P{2.5cm}|P{2.5cm}|}
         \hline
        Modules & Static & Static \& Emulation & All \\
         \hline
        Original set & 0 & 5 & 61 \\
        \hline
        Adversarial set & 1515 & 236 & 80 \\
        \hline
        Evasion rate, $\epsilon$ & 28.06\% & 4.28\% & 0.35\% \\
        \hline
    \end{tabular}
    \label{tab:evasive_ember}
\end{table*}

Our experiments show that not all adversarial samples are functional after modifications. We observe that on average about 40\% of binary files after section injections are not functional, with detailed statistics reported in Table \ref{tab:adv_emulation_stats}. Based on manual analysis, we can conclude that those represent packed malware samples, and we assume that section injection interferes with the unpacking routine. We perform the final analysis only on functional adversarial malware, considering broken samples as unsuccessful and discarding from further statistics. Additionally, some of the GAMMA executions fail with \texttt{AttributeError, FitnessMin} exception, which are subject of \texttt{secml\_malware} library limitations, therefore the number of total samples in Table \ref{tab:adv_emulation_stats} is variable.

The focus of adversarial robustness tests is to evaluate whether simultaneous utilization of static, dynamic, and contextual parameters of executed malware will provide the necessary epistemic background to lessen the effects of an adversarial attack. We evaluate an absolute number of evasive samples, evasion rates, and detection accuracy based on different configurations of enabled modules in a hybrid model. 

We use the \textit{evasion rate} to illustrate attack's efficiency ratio against a specific model setup. If the effect of the GAMMA attack on sample is denoted as $\delta$, true label of sample as $y$, then: 
$$e' \in A: \psi(\phi(e+\delta)) \neq y,$$
$$e \in A:\psi(\phi(e)) \neq y,$$
with $e'$ representing a number of evasive samples in an adversarial set, and $e$ evasive samples in original set before execution of the attack. $\phi$ represents early fusion pass, $\psi$ - late fusion pass as denoted in Eq. \ref{eq:late_fusion}. The evasion rate is calculated as follows: 
$$\epsilon = \frac{\Delta e}{|A|} = \frac{e' - e}{|A|},$$ 
where $|A|$ is total number of functional samples produced by attack, and $\Delta e$ represents of evasive samples added by the attack. 

We acknowledge the high efficiency of the GAMMA attack against static classifiers. We see that detection rates of static module drop from $100\%$ to $72\%$. Quantitative statistics for an attack against the static model with 15 injected sections are reported in Table \ref{tab:evasive_ember}. The adversarial algorithm is capable of producing \textbf{1515} evasive executables for the static model based on ember feature engineering. By adding emulation-produced API calls to a hybrid model, evasive sample counts drop to \textbf{236}. Contextual information in form of file paths decreases counts further to \textbf{80}.

GAMMA produces adversarial samples only from already detected malware by the target model. Therefore, Table \ref{tab:evasive_ember} reports zero evasive samples for static model in original set. However, adding modules to a hybrid solution shifts the decision boundary. Thus, we observe that some of the samples from the original set were evasive for the hybrid model before GAMMA manipulations. For instance, $5$ samples were marked as benign when static and API call models are employed together, and $61$ by adding a filepath model to the final heuristic.

The evasion rate shows that in case of hybrid model, GAMMA's contribution to evasion abilities is even less significant since some were evasive before adversarial manipulations. Thus, adding emulation-based analysis drops the evasion rate from $28.06\%$ to $4.28\%$, file path module reduces it further to $0.35\%$, practically eliminating the attack's effect.


\section{Conclusion and Future Work}\label{sec:conclusions}

This work presents a hybrid machine learning architecture that employs the Windows portable executable's static, behavioral, and contextual properties. We performed behavioral analysis on large corpora of executables collected from in-the-wild systems and labeled by a professional threat intelligence team. 

We have shown that ML algorithms benefit from hybrid analysis, yielding improved performance, especially under low false-positive requirements. We indeed report exceptional performance by the current state-of-the-art malware modeling scheme based on Ember feature vector \cite{anderson2018ember}, which reports detection rates significantly higher than filepath or emulation models individually, as seen in Figure \ref{fig:rate_heatmap}. However, combining the Ember model with either filepath or emulation, or both models notoriously improve detection capabilities, under some circumstances like low false positive requirements by almost $30\%$.

Additionally, we report that a hybrid solution can detect a malevolent sample even if none of the individual components express enough confidence to classify input as malicious. For instance, with FPR of one misclassified case in $10^5$, individual filepath and emulation models detect only $33.31\%$ and $10.24\%$ samples. A combination of them produces a $77.36\%$ detection rate, increasing the detection abilities by more than $40\%$ if both models were used together but independently.

This observation holds across in-sample validation and out-of-sample test sets, collected from divergent systems and in different time frames, allowing us to conclude that this is a property of a hybrid solution rather than an artifact of a specific dataset. We conclude that the meta-model can learn patterns typical to malicious samples out of representation combinations produced by different analysis techniques. Furthermore, this conclusion is supported by the dataset size, significantly larger than in related works on behavioral malware analysis.

Empirical observations show that the GAMMA attack has low efficiency against hybrid heuristic that incorporates static, dynamic, and contextual features of PE. We assume this extrapolates to other adversarial techniques targeted at altering PE properties that modify the feature space of static detectors. These attacks do not affect PE properties evaluated by emulation and file path module. A successful attack against such a hybrid system should incorporate multipurpose action space that allows altering static, dynamic, and contextual properties.

We suggest that the positive traits of dynamic and contextual analysis can be extended further. While we represent PE behavior on a system with an API call sequence, not all executable functionality is expressed through API calls. Additional visibility sources might be crucial to minimize ambiguity in the model decision heuristic, and we argue that extending the modularity of composite solutions is a promising research direction. Most of the emulation telemetry was omitted from our analysis. 
We publicly release emulated reports of $108204$ samples and expect further work in this direction. File system and registry modifications, network connections, and memory allocations may provide crucial information for detection. The architecture of our solution allows us to extend the modularity of the decision heuristic with minimal effort by retraining only the parameters of a meta-model.

\balance


\begin{acks}
We express a deep gratitude for Marlon Tobaben, Antti Honkela and Maria Regaki for contributing towards final version of this publication.
\end{acks}

\bibliographystyle{ACM-Reference-Format}
\bibliography{sample-base}


\end{document}